\documentclass[letterpaper,12pt]{article}
\pdfoutput=1

\usepackage{jheppub}
\usepackage{amsmath}
\usepackage{graphicx}
\usepackage{subfig}
\usepackage{cancel}
\usepackage[dvipsnames]{xcolor}
\usepackage{color}
\usepackage{mathrsfs}

\usepackage{amssymb}
\usepackage{epstopdf}
\def\({\left(}
\def\){\right)}
\def\[{\left[}
\def\]{\right]}

\newpage

\title{Does horizon entropy satisfy a Quantum Null Energy Conjecture?}

\author{Zicao Fu}
\author{and Donald Marolf}

\affiliation{Department of Physics, University of California, Santa Barbara, CA 93106, USA}

\emailAdd{zicaofu@physics.ucsb.edu}
\emailAdd{marolf@physics.ucsb.edu}

\abstract{A modern version of the idea that the area of event horizons gives $4G$ times an entropy is the Hubeny-Rangamani Causal Holographic Information (CHI) proposal for holographic field theories.  Given a region $R$ of a holographic QFTs, CHI computes $A/4G$ on a certain cut of an event horizon in the gravitational dual.  The result is naturally interpreted as a coarse-grained entropy for the QFT.  CHI is known to be finitely greater than the fine-grained Hubeny-Rangamani-Takayanagi (HRT) entropy when $\partial R$ lies on a Killing horizon of the QFT spacetime, and in this context satisfies other non-trivial properties expected of an entropy.  Here we present evidence that it also satisfies the quantum null energy condition (QNEC), which bounds the second derivative of the entropy of a quantum field theory on one side of a non-expanding null surface by the flux of stress-energy across the surface.
In particular, we show CHI to satisfy the QNEC in 1+1 holographic CFTs when evaluated in states dual to conical defects in AdS$_3$.  This surprising result further supports the idea that CHI defines a useful notion of coarse-grained holographic entropy, and suggests unprecedented bounds on the rate at which bulk horizon generators emerge from a caustic.  To supplement our motivation, we include an appendix deriving a corresponding coarse-grained generalized second law for 1+1 holographic CFTs perturbatively coupled to dilaton gravity.}

\begin{document}
\maketitle

\section{Introduction}
\label{sec:Introduction}

The Hawking area theorem \cite{Hawking:1973uf} states that the total event horizon area cannot decrease.  As a result, it has long been conjectured \cite{Bekenstein:1973ur} that this event horizon area measures some notion of entropy, and that --  at least at the semiclassical level -- quantum field theories coupled to gravity there will be a non-decreasing quantity of the form
\begin{equation}
\label{Sgen}
S_{gen} = S_{BH} + S_{QFT},
 \end{equation}
given by the sum of the Bekenstein-Hawking entropy $S_{BH} = \frac{A}{4G_d}$ of black holes and the entropy $S_{QFT}$ of quantum field theories (QFT$_d$'s) outside.  This conjecture is known as the generalized second law (GSL).  Here $d$ is the spacetime dimension and $G_d$ is Newton's gravitational constant.

The Hubeny-Rangamani Causal Holographic Information (CHI) construction \cite{Hubeny:2012wa} attempts to connect this idea with gauge/gravity duality by interpreting the area of certain cuts of event horizons in the higher-dimensional gravitational dual as measuring some sort of entropy for associated regions $\Sigma$ in the $d$-dimensional holographic QFT$_d$.  As we review in section \ref{sec:setting}, CHI is $\frac{1}{4G_{bulk}}$ times the area of a the appropriate cut; note that this $G_{bulk}$ is unrelated to the any Newton constant $G_d$ that might be used to couple our holographic QFT$_d$ to dynamical gravity.   Because CHI is bounded below \cite{Hubeny:2012wa,Maximin} by the Hubeny-Rangamani-Takayanagi (HRT) entropy $S_{HRT}$ defined by extremal surfaces \cite{HRT}, it is natural to interpret CHI as a coarse-grained entropy; see \cite{CHI,KellyWall} for specific proposals of possible corresponding field theory coarse-graining procedures.

Further support for this idea comes from \cite{coarse-grainedGSL}, which considered perturbatively coupling (the universal sector of) a holographic CFT$_d$ to dynamical gravity with Newton constant $G_d$.  In particular, \cite{coarse-grainedGSL} showed the universal sector of $d > 2$ holographic theories to satisfy a next-to-leading-order $G_d \to 0$ version of the GSL with $S_{QFT}$ given by a renormalized version of CHI.  Here it is important to note that \cite{coarse-grainedGSL} renormalized CHI using the same counterterms as are required for corresponding HRT entropies $S_{HRT}$.   In general, renormalizing CHI in this way would still leave a divergence \cite{CHI}. But, at least when the state approaches equilibrium as one moves forward along a (conformal) Killing horizon, the derivation in \cite{coarse-grainedGSL} showed the result to give a finite renormalized coarse-grained entropy $S^{coarse}_{QFT}$ for regions $\Sigma$ of the QFT$_d$ whose boundary $\partial \Sigma$ lies on this horizon.

Below, we provide evidence that the entropy defined by CHI satisfies another non-trivial inequality -- the Quantum Null Energy Condition (QNEC)  -- that might be expect to hold for useful notions of entropy in a QFT$_d$.  The QNEC for von Neumann (fine-grained) entropy was conjectured in \cite{Bousso:2015mna} to hold for any QFT$_d$ as a result of studying the implications of taking the $G_d\to 0$ limit of either the GSL or possible covariant entropy bounds \cite{Bousso:1999xy,Bousso:1999cb,Flanagan:1999jp} on the flux of entropy through null surfaces.  We will write the QNEC in the form
\begin{equation}
\label{QNEC}
\int_{\partial \Sigma} T_{\alpha \beta} k^\alpha k^\beta \ge \frac{1}{2\pi} \frac{d^2}{d\lambda^2} S_{QFT},
\end{equation}
in terms of the renormalized stress tensor $T_{\alpha \beta}$ and renormalized\footnote{The original conjecture of \cite{Bousso:2015mna} was stated in terms of the un-renormalized entropy so that  our counterterms do not appear.  But in the contexts studied in \cite{Bousso:2015mna,Holofine-grainedGSL} the counterterms are independent of $\lambda$ and are annihilated by derivatives, so \eqref{QNEC} is unaffected by our including such terms.  In more generality the counterterms contribute, and without them the right hand side of \eqref{QNEC} can have divergences of either sign; see \cite{Marolf:2016dob} for related discussion.  It is thus clear that a general QNEC must involve renormalized entropy, so we include couter-terms as a ``friendly ammendment'' to the conjecture of \cite{Bousso:2015mna}. } von Neumann entropy
\begin{equation}
S_{QFT}(\Sigma) = - Tr(\rho \ln \rho) + counterterms
 \end{equation}
of our QFT$_d$ in a region $\Sigma$ of a Cauchy surface in the $d$-dimensional background spacetime.  The counter-terms in both $T_{\alpha \beta}$ and $S_{QFT}$ are to be defined by the usual covariant renormalization of the partition function, taking $T_{\alpha \beta}$ and $S_{QFT}$ to be given respectively by variations with respect to the background metric and use of the replica trick.  The effects of any remaining scheme dependence on the QNEC will be discussed in future work \cite{SchemeDep}.

In \eqref{QNEC}, the derivatives on the right-hand-side are defined by considering the ingoing null congruence launched orthogonally from $\Sigma$ labelled by an affine parameter $\lambda$, with $k^\alpha\partial_\alpha = \frac{d}{d\lambda}$ and $\theta$ being respectively the associated  tangents and expansion.  This fine-grained QNEC was conjectured \cite{Bousso:2015mna} to hold when the quantity $\theta k^\alpha$ vanishes at all points of $\partial \Sigma$; i.e., when the congruence is allowed to advance at a point $p$ of $\partial \Sigma$ only if $\theta$ vanishes there.   The QNEC has been proven \cite{Bousso:2015mna} for superrenormalizeable bosonic fields when $\partial \Sigma$ lies on a Killing horizon, and (building on the results of \cite{Maximin}) for the universal sector of holographic conformal field theories (CFTs) on flat spacetime \cite{Holofine-grainedGSL} using the Hubeny-Rangamani-Takayanagi (HRT) proposal to compute $S_{QFT}$ via extremal surfaces. We understand that this proof can also be generalized to Killing horizons of more general spacetimes \cite{private}.  We also mention the related work \cite{Faulkner:2016mzt}, which establishes is for general relativistic field theories (and, in particular, more broadly than the QNEC has been shown to hold) the a related but logically-weaker condition known as the averaged null energy condition (ANEC)\footnote{The ANEC is just an integral of $T_{\alpha \beta} k^\alpha k^\beta$ and can be derived by integrating the QNEC over $\lambda$ and enforcing boundary conditions that require $S' \rightarrow 0$ in the far past and future.}.

Putting these together, by the same reasoning as in \cite{Bousso:2015mna} the $G\rightarrow 0$ limit of the GSL \cite{coarse-grainedGSL} for CHI suggests that CHI may also satisfy \eqref{QNEC}. Intriguingly, the derivation of the CHI coarse-grained GSL in \cite{coarse-grainedGSL} in fact showed that certain contributions to $\frac{S_{QFT}^{coarse}}{d\lambda}$ exactly saturate \eqref{QNEC}.
Roughly speaking, these were the contributions  computed in the bulk dual from the flux $F$ of bulk horizon generators through the bulk conformal boundary, or more properly by taking the  $z_0 \to 0$ limit of the flux through a regulating surface at constant Fefferman-Graham coordinate $z=z_0$.  This part of the change was denoted $\frac{d}{d\lambda}S_{QFT, \ adiabatic}$ in \cite{coarse-grainedGSL}, but we write it as simply $\frac{1}{4G_{bulk} }F$ below in terms of the Newton constant $G_{bulk}$ of the dual bulk description.

The remaining pieces of $\frac{dS_{QFT}^{coarse}}{d\lambda}$ come from the growth of area along individual bulk generators and from the emergence of horizon generators from caustics. In \cite{coarse-grainedGSL} their sum was denoted $\frac{d}{d\lambda}S_{QFT, \ non-dec}$, but we will call this sum  $\frac{1}{4G_{bulk} }\frac{dA_{\text{Hawking}}}{d\lambda}$ as it contains precisely the (non-decreasing) contributions that are controlled by the Hawking area theorem \cite{Hawking:1971tu} in the bulk.   Note that there is no corresponding division of the undifferentiated quantity $S_{QFT}^{coarse}$ into parts associated with $F$ and $\frac{dA_{\text{Hawking}}}{d\lambda}$; the latter quantities are fundamentally associated with rates of change. The coarse-grained GSL of \cite{coarse-grainedGSL} followed quickly from the above-mentioned saturation of \eqref{QNEC} by $\frac{F}{4G_{bulk}}$ and the fact that the bulk area theorem requires $\frac{dA_{\text{Hawking}}}{d\lambda} \ge 0$.

Let us now return to the possibility of a CHI coarse-grained QNEC. We emphasize that this is not logically required by either the conjecture of \cite{Hubeny:2012wa} that CHI represents a coarse-grained entropy or by the fine-grained QNEC conjecture of \cite{Bousso:2015mna}, but instead represents a further conjecture in itself. In any case, since
$\frac{1}{4G_{bulk}}F$ already saturates \eqref{QNEC},  satisfaction of \eqref{QNEC} by CHI requires
\begin{equation}
\label{QNECirr}
\frac{d^2}{d\lambda^2}A_{\text{Hawking}} \le 0.
\end{equation}
As \eqref{QNECirr} resembles the Raychaudhuri equation satisfied by null geodesics, the reader might not be surprised if \eqref{QNECirr} turns out to be satisfied by the contributions from the growth of area along individual bulk generators.  This is especially true for $d=2$ (i.e., in three bulk dimensions) where the Raychaudhri equation and null convergence condition together imply the area element along each generator to have negative second derivative computed with respect to the bulk affine parameter.  But  the $\lambda$ of \eqref{QNECirr} is specified by the CHI prescription of \cite{Hubeny:2012wa} and need not be affine in the bulk.  Furthermore, we have found no precedent in the literature for a bound of any sort on the rate at which bulk horizon generators emerge from a caustic.  Nevertheless, our results below support the conjecture that \eqref{QNECirr} may in fact hold.

Lacking any general tools to study \eqref{QNECirr}, one is naturally led to testing the possibility of a coarse-grained QNEC through various examples. For simplicity, we test \eqref{QNECirr} below for holographic CFTs with $d=2$, and thus in bulk spacetimes asymptotic to AdS$_3$. The expansion of any null geodesic vanishes trivially in any 2d spacetime, so taking $\Sigma$ to range over all intervals we see that our coarse-grained version of \eqref{QNEC}  reduces to two independent conditions corresponding to left- and right-moving null geodesics at each point of the spacetime.

We will focus on the constraint that \eqref{QNECirr} imposes on the rate at which bulk horizon generators emerge from caustics.  In particular, we study CFT states on 1+1 Minkowski space that can be obtained from bulk duals given by point particles in otherwise-empty AdS$_3$.  As explained in section \ref{sec:setting}, this means that we study Rindler horizons in conical defect spacetimes.  The bulk expansion $\theta_{bulk}$ then vanishes along each generator and the only contributions to \eqref{QNECirr} come from caustics.  Lest the reader be concerned that part of our motivation and reasoning above relied on the $d > 2$ arguments of \cite{coarse-grainedGSL}, we extend those results to $d=2$ in appendix \ref{sec:Append_2dGSL}.  This appendix also goes beyond the universal sector and allows QFTs with general scalar and metric sources.

There is, however, a further important point to consider for $d=2$.  This point is associated with the fact that the conformal group becomes infinite dimensional. As noted in \cite{Wall:2011kb}, the relation \eqref{QNEC} does not transform homogeneously under this symmetry.  A much better behaved relation would take the form
\begin{equation}
\label{2dQNEC}
 Q:= T_{\alpha \beta} k^\alpha k^\beta - \frac{1}{2\pi}\left[  \frac{d^2}{d\lambda^2} S_{QFT} + \frac{6}{c} \left( \frac{d}{d\lambda} S_{QFT} \right)^2 \right] \ge 0,
\end{equation}
where $c = \frac{3\ell}{2G_{bulk}}$ is Virasoro the central charge determined by $G_{bulk}$ and the bulk AdS scale $\ell$.   Making the change of 2d line element $d\tilde s^2 =  \Omega^2 ds^2$, the quantity $\tilde Q$ defined with respect to $d\tilde s^2$ is related to the original $Q$ defined relative to $ds^2$ by simply $\tilde Q = \Omega^{-2} Q$. As a result, a failure of \eqref{2dQNEC} requires \eqref{QNEC} to fail in the conformal frame where $\frac{d}{d\lambda}S_{QFT}=0$ at $p$.   This conformal frame always exists due to the conformal anomaly.   And since \eqref{2dQNEC} also clearly implies \eqref{QNEC}, the two statements are equivalent in 2d QFTs so long as both are imposed in all conformal frames.  But the advantage of studying \eqref{2dQNEC} is that, due to its simple transformation properties, checking that \eqref{2dQNEC} holds in any given frame (say, one chosen to may the calculation easy) then implies both \eqref{2dQNEC} and \eqref{QNEC} to hold in all conformal frames.   While the above reasoning may appear to hold only for CFTs, it can be generalized by requiring \eqref{QNEC}, \eqref{2dQNEC} to hold for theories with spacetime-dependent couplings.

Returning to our motivations, it is important to note that the fine-grained version of \eqref{2dQNEC} was derived directly for $d=2$ holographic CFTs in \cite{Holofine-grainedGSL}.  Furthermore, in appendix \ref{sec:Append_2dGSL} we show that for $d=2$ holographic QFTs the quantity $\frac{1}{4G_{bulk}} F$ in fact saturates \eqref{2dQNEC} instead of \eqref{QNEC}; recall that \cite{coarse-grainedGSL} studied only $d\ge 3$.  Due to the quadratic term in \eqref{2dQNEC}, the analogue of \eqref{QNECirr} generally takes the form
\begin{equation}
\label{eq:2dcoarse-grainedQNEC}
\frac{d^2}{d\lambda ^2}A_\text{Hawking}+\frac{1}{\ell}\left(\frac{d}{d\lambda }A_\text{Hawking}\right)^2+\frac{2F }{\ell} \frac{d}{d\lambda }A_\text{Hawking}\leq 0.
\end{equation}
However,  section \ref{sec:setting} identifies a frame in which $F$ vanishes so that \eqref{eq:2dcoarse-grainedQNEC} then depends only on $\frac{dA_{Hawking}}{d\lambda}$.

We begin in section \ref{sec:setting} below by explaining the particular states to be studied in our holographic CFT, along with a variety of conformal frames that will prove useful. This sets the stage for section \ref{sec:testing} to show that points $p$ on a certain null plane indeed respect \eqref{2dQNEC} and \eqref{eq:2dcoarse-grainedQNEC}; the generalization to all $p$ is relegated to appendix \ref{sec:Append_GeneralHorizons}.   We close with a discussion of open issues in section \ref{sec:discussion}.  The above-advertised $d=2$ proof of the coarse-grained holographic GSL appears in appendix \ref{sec:Append_2dGSL}, while appendix \ref{AdS3rev} contains a brief review of coordinates and horizons in empty AdS$_3$ to assist readers wishing to put certain equations from the main text into the proper context.

\section{Setting the Stage}
\label{sec:setting}

\begin{figure}[t]
\centerline{
\includegraphics[width=0.5\textwidth]{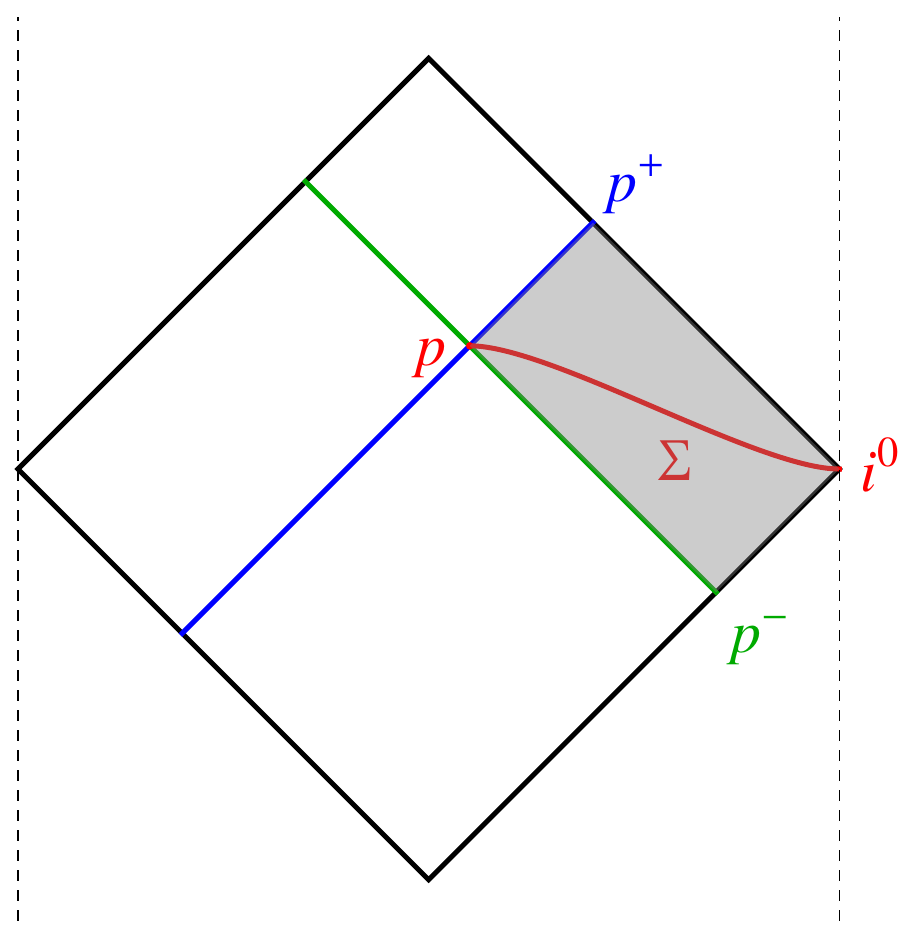}
}
\caption{A conformal diagram of our 1+1 Minkowski space showing $p$, $p^+$, $p^-$ and $i^0$ with the associated past (blue) and future (green) Rindler horizons and a partial Cauchy surface $\Sigma$ extending from $p$ to the right to $i^0$. The domain of dependence $D(\Sigma) = I^-(p^+) \cup I^+(p^-)$ (shaded) is the right Rindler wedge.   The dashed vertical lines define a strip.  Identifying them conformally maps our Minkowski space into $S^1 \times {\mathbb R}$.  }
\label{fig:2dCFT}
\end{figure}

Due to the simple conformal transformation properties of the proposed 2d QNEC \eqref{2dQNEC} and the fact that all 1+1 spacetimes are conformally flat, it suffices to test our coarse-grained QNEC at some point $p$ in $1+1$ Minkowski space.  Without loss of generality, we take our coarse-grained holographic entropy $S_{QFT}^{coarse}$ to refer to the region $\Sigma$ to the right of $p$.  As specified in \cite{Hubeny:2012wa}, the CHI of $\Sigma$ is determined by the domain of dependence $D(\Sigma)$ in our Minkowski space, which is precisely the right Rindler wedge $\Delta \bar x > |\Delta \tau|$ for $\Delta \bar x = \bar x- \bar x(p)$, $\Delta \bar \tau = \bar \tau - \bar \tau(p)$ and $\bar x,\bar \tau$ the usual Minkowski inertial coordinates in some frame.  As a result, as shown in figure \ref{fig:2dCFT}, we may also determine $D(\Sigma)$ from the points $p_\pm$ where the past and future Rindler horizons through $p$ reach past and future null infinity.  Indeed, $D(\Sigma) = I^-(p_+) \cap I^+(p_-)$, where $I^\pm$ denote the regions in our 1+1 Minkowski space that lie in  the (timelike) past and future of the indicated points.  Since we study conformal field theories, it is useful to recall  that 1+1 Minkowski space is conformally equivalent to a region of the $1+1$ cylinder $ S^1 \times {\mathbb R}$, and in particular that $p_\pm$ thus map to finite points.  CFT states that approach the vacuum sufficiently quickly near the point $i^0$ at spacelike infinity of our Minkowski space will map to smooth states on this cylinder.

Our coarse-grained entropy will be a renormalized version of the causal holographic information  defined by the above region $\Sigma$ for CFT states dual to asymptotically AdS$_3$ bulk geometries.  In an appropriate conformal frame the conformal boundary will again be a cylinder and will have points corresponding to $p^\pm$ and also to $i^0$.  The corresponding CHI is then computed by finding the past of $p^+$ in the bulk (which we call $I^-_{bulk}(p^+)$) and the corresponding bulk future $I^+_{bulk}(p^-)$ of $p^-$.  In particular, defining $H^\pm_{bulk}(p)$ to be the boundary of $I^\mp_{bulk}(p^\pm)$, we have
\begin{equation}
\label{2dCHI}
S_{QFT}^{coarse} = \lim_{z_0 \to 0} \left[\frac{{\rm Length}_{z>z_0} \  {\rm of} \ H^+_{bulk}(p) \cap H^-_{bulk}(p) }{4G_{bulk}} +  \frac{\ell}{4G_{bulk}}  \ln (2z_0/\ell) \right],
\end{equation}
where the first term refers to the length of the part of $H^+_{bulk}(p) \cap H^-_{bulk}(p) $ in the region $z > z_0$ as specified by some bulk Fefferman-Graham coordinate $z$ associated with the given conformal frame on the boundary.

Up to a conventional finite piece,
the second term in \eqref{2dCHI} is  a standard version of the $d=3$ HRT counterterm; see e.g. \cite{Graham:1999pm,RT}.  Despite our inclusion of this counter term at $p$, the quantity \eqref{2dCHI} still diverges since $H^+_{bulk}(p) \cap H^-_{bulk}(p)$   will also approach the point $i^0$ on the AdS$_3$ boundary.  But we will hold $i^0$ fixed when computing $ \frac{A_{\text{Hawking}}}{d\lambda}$ and the flux $F$ of horizon length through the boundary, so divergences or counterterms at $i^0$ cannot contribute.
Below, we refer to $H_{bulk}^\pm(p)$ as the past and future bulk Rindler horizons defined by the point $p$ on the boundary.

We will consider CFT states $|\psi \rangle$
dual to asymptotically AdS$_3$ spacetimes of the form
\begin{equation}
\label{eq:pointmass}
d\bar{s}^2=-\left(\frac{\bar{r}^2}{\ell ^2}-M \right)d\bar{t}^2+\frac{1}{\frac{\bar{r}^2}{\ell ^2}-M}d\bar{r}^2+\bar{r}^2d\bar{\phi }^2.
\end{equation}
For $M=-1$ this is just empty $\mathrm{AdS_3}$ in global coordinates, while for $M \ge 0$ it represents a BTZ black hole of ``mass'' $M$ in the conventions of \cite{Banados:1992gq}.   But for $-1<M<0$ we write $M=-\alpha^2$ and our spacetimes may be interpreted as being sourced by a point mass $m$ at the origin with $\alpha=1-4G_{bulk}m$ in terms of the AdS$_3$ Newton constant $G_{bulk}$.  Indeed, for such $M$ \eqref{eq:pointmass} is related to the point mass spacetimes originally described in \cite{Deser:1983nh} by a simple coordinate transformation.
 The corresponding CFT states $|\psi \rangle$ are related via the state-operator correspondence to operators with dimensions $\Delta = \gamma c$ for $\gamma \in [0,1/12]$ in terms of the CFT central charge $c$.

Pulling out a conformal factor $\frac{\bar r^2}{\ell^2}$, the metrics \eqref{eq:pointmass} are naturally associated with a conformal frame in which the boundary metric is $S^1 \times {\mathbb R}$ and the $S^1$ has radius $\ell$.  As a result, we can take the inertial coordinates $\bar x, \bar \tau$ of the CFT's original 1+1 Minkowski space to be related to $(\bar t, \bar \phi)$ in an essentially standard way -- though for reasons explained in section \ref{sec:testing} below we find it useful to shift to orgin $(\bar \tau ,\bar x) = (0,0)$ to lie at $(\bar t, \bar \phi) = \left(0, \frac{\pi}{2}\left(\frac{1}{\alpha}-1\right)\right)$ so that $i^0$ lies at $(\bar t, \bar \phi) = (0, \frac{\pi}{2\alpha})$.  In particular, we take
\begin{equation}
\label{eq:bargtoP}
\tan \frac{\bar t}{\ell} = \frac{2 \ell \bar \tau}{\ell^2 + \bar x^2 -\bar \tau^2}, \ \
\tan \left( \bar \phi - \frac{\pi}{2} \left(\frac{1}{\alpha}-1\right)\right)   = \frac{-\ell^2 + \bar x^2 -\bar \tau^2}{2 \ell \bar x}.
\end{equation}
As reviewed in appendix \ref{AdS3rev}, one may think of this as the boundary limit of the relation between global and certain Poincar\'e coordinates on empty AdS$_3$.  This conformal frame is also associated with some Fefferman-Graham coordinate $\bar z$, though we will have no need for its explicit form. We refer to \eqref{eq:pointmass} as the global description of our spacetime (and the corresponding boundary as the global conformal frame), and we similarly refer to $(\bar \tau,\bar x)$ as the Poincar\'e frame.

As noted at the end of the introduction, \eqref{2dQNEC} simplifies in a conformal frame where $T_{\alpha \beta}k^\alpha k^\beta=0$.  We may find such a conformal frame using the fact that
as noted in \cite{Deser:1983nh,Banados:1992gq}, the point mass spacetimes are equivalent to conical defects in empty global AdS$_3$.  In particular, writing $\bar{t}=\frac{t}{\alpha },\bar{r}=\alpha r,\bar{\phi }=\frac{\phi }{\alpha }$, the metric takes the form
\begin{equation}
\label{eq:AdS3}
ds^2_{GmD}=-\left(1+\frac{r^2}{\ell ^2}\right)dt^2+\frac{1}{1+\frac{r^2}{\ell ^2}}dr^2+r^2d\phi ^2.
\end{equation}
This coincides with empty AdS$_3$ in global coordinates, except that in \eqref{eq:AdS3} the coordinate $\phi$ has period $2\pi \alpha$.  As a result, we refer to this as the ``global minus defect'' (GmD) description of our bulk spacetime.  We denote the associated defect angle by $\Gamma = 2\pi(1-\alpha)$. It will be convenient below to take $\phi$ to range over the symmetric interval $[-\pi + \Gamma/2, \pi- \Gamma/2]$. We refer to the missing values $\phi \in [-\pi, -\pi + \Gamma/2] \cup [\pi - \Gamma/2 , \pi]$ as ``the defect'' below.  Due to the defect, pulling out a factor of $\frac{r^2}{\ell^2}$ shows that \eqref{eq:AdS3} is naturally associated with a conformal frame in which the boundary metric is $S^1 \times {\mathbb R}$ where the $S^1$ has radius $\ell(1-\Gamma/2\pi)$.

\begin{figure}[t]
\centerline{
\includegraphics[width=0.5\textwidth]{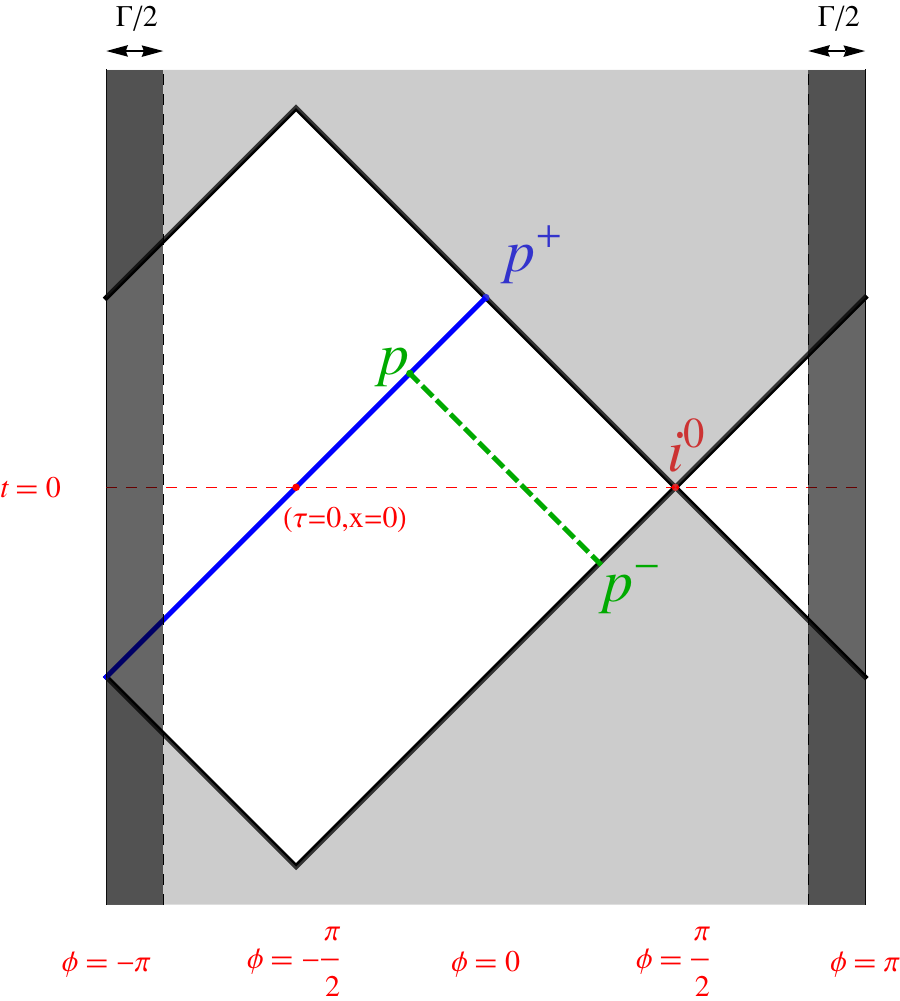}
}
\caption{The conformal boundary in our Global minus Defect (GmD) conformal frame for the case of defect angle $\Gamma < \pi$.  The unshaded region is that associated with the boundary in the Poincar\'e minus Defect (Pmd) frame.  Since $\Gamma < \pi$, the points $p^+$, $p^-$, and $i^0$ cannot enter the defect.  However, the point $p$ still reaches the defect at $t=- \pi/2 + \Gamma/2 \le 0$. We have placed $p^+$ at $\phi =0$ as this will be the case emphasized in section \ref{sec:testing}.}
\label{fig:GmDPmD}
\end{figure}

The GmD frame clearly has the same boundary stress tensor as empty AdS$_3$.  So $T_{\alpha \beta}k^\alpha k^\beta=0$ still does not vanish, but only one step remains to construct a frame where it does.  Away from the defecet we may use our GmD coordinates $(t, \phi, r)$ to introduce new  `Poincar\'e minus Defect' (PmD) coordinates $(x,\tau,z)$ by through the relations that would map between global and Poincar\'e coordinates in empty AdS$_3$.  Choosing $i^0$ of the PmD frame to coincide with $i^0$ of the original $(\bar \tau, \bar x)$ Poincar\'e frame introduced above (and thus to lie at $(t,\phi) = (0,\pi)$), we have
\begin{align}
\label{eq:tant}
& \tan \frac{t}{\ell }=\frac{2\ell \tau }{z^2+\ell ^2+x^2-\tau ^2}, \\
\label{eq:tanphi}
& \tan \phi =\frac{z^2-\ell ^2+x^2-\tau ^2}{2\ell x}, \\
\label{eq:r}
& r^2=\frac{\ell ^2x^2}{z^2}+\frac{\left(z^2-\ell ^2+x^2-\tau ^2\right)^2}{4z^2},
\end{align}
which also locates the PmD origin $(\tau, x,z) =(0,0,0)$ at $(t,\phi,r) = (0, - \pi/2, +\infty)$ as shown in figure \ref{fig:GmDPmD}.   Just as in the Poincar\'e patch of empty AdS$_3$ the bulk metric must then become
\begin{equation}
\label{eq:poincaremetric}
ds^2_{PmD}=\frac{\ell ^2}{z^2}\left(dz^2-d\tau^2+dx^2\right),
\end{equation}
though with coordinate ranges defined to exclude the above defect.   While the detailed specification of the defect $\phi \in [-\pi, -\pi + \Gamma/2] \cup [\pi - \Gamma/2 , \pi]$ is complicated in the PmD description, this feature will cause no problems.

Away from the defect the computation of the PmD boundary stress tensor is trivial; \eqref{eq:poincaremetric} gives $T_{\alpha \beta} =0$.  We recall that appendix \ref{sec:Append_2dGSL} shows $\frac{F}{4 G_{bulk}}$ to saturate \eqref{2dQNEC}, so $T_{\alpha \beta}=0$ suggests that $F$ vanishes as well\footnote{Saturating \eqref{2dQNEC} gives an ODE which requires any non-zero $\frac{d}{d\lambda} S_{QFT, \ adiabatic}$ to diverge at some finite affine parameter.  But this is not necessarily a contradiction as the PmD conformal frame is singular at the edges of the defect and these edges are reached at finite PmD affine parameter $\lambda$.  This singularity is associated with the fact that
 the conformal factor relating PmD to GmD is not continuous across the identifications.  If this feature seems unpalatable, the reader is free to replace PmD with a smooth conformal frame that is identical except in some small neighborhood of the defect $\phi \in [-\pi, -\pi + \Gamma/2] \cup [\pi - \Gamma/2 , \pi]$.  Taking the limit in which the neighborhood shrinks to zero size is equivalent to using the singular PmD frame, though at any finite step in this limit $T_{\alpha \beta}$ will remain non-zero -- and in fact will become large -- in a small region associated with the defect.}.  Having chosen $i^0$ in the PmD and original Poincar\'e frames to coincide, $p^+$ will also lie on left-moving future null infinity in the PmD frame.  As a result, the future Rindler horizon (the past light cone of $p^+$ in the bulk) takes the form $x =\tau + constant$ in the PmD frame, with each generator having constant $z$. The flux of horizon area thus vanishes through any $z=z_0$ regulating surface.  Since the counterterm in \eqref{2dCHI} is also independent of $(x,\tau)$, we find $F=0$ in the PmD frame as claimed.

 As explained in the introduction, this in turn means that in the PmD frame our QNEC quantity $Q_{PmD}$ is determined entirely by $\frac{dA_{\text{Hawking}}}{d\lambda}$.  In principle this term involves both the local expansion along each generator of $H^+_{bulk}$ and the rate $dL_c/d\lambda$ at which horizon length emerges from the caustic as measured by a PmD affine parameter at $p$; e.g. $\lambda = \tau(p)$.  But since the bulk past light cone of $p^+$  is locally just a Rindler horizon in empty AdS$_3$, all expansions vanish.  Thus we have simply
 \begin{equation}
 \label{QPMD}
 Q_{PmD} =- \frac{1}{8\pi G_{bulk}} \left[\frac{d^2L_c}{d\lambda^2} + \frac{1}{\ell} \left(\frac{d L_c}{d\lambda} \right)^2 \right].
 \end{equation}
 Note that $dL_c/d\lambda$ is a well-defined finite quantity that is free of anomalies and whose transformation between conformal frames is determined by the reparameterization $\lambda(\tilde \lambda)$ dictated by the associated affine parameters $\lambda, \tilde \lambda$.  To verify \eqref{2dQNEC} in all frames, it thus suffices to check that $Q_{PmD} \ge 0$.

While the definition of our coarse-grained entropy is fundamentally symmetric in $p^+$ and $p^-$, we have thus far focussed on $p^+$.  Indeed, the reader will note that the only role for $p^-$ in our approach is in determining the parameter $\lambda$.  This limited role is due to the above-mentioned vanishing expansion along each bulk horizon generator so that the area of any cut of this horizon depends only on whether and where the cut intersects the caustic.  For the same reason we have had little to say about the point $i^0$; the area of the CHI surface near $i^0$ cannot change except where the caustic reaches the conformal boundary of the bulk.

\section{Testing the coarse-grained QNEC}
\label{sec:testing}

We are now ready to test our coarse-grained QNEC. We choose to proceed in two steps.  First, we compute $dL_c/dt$, which is just the rate at which length emerges from the caustic as measured by the GmD coordinate $t$.  Later we will find $dt/d\lambda$.  For simplicity, we focus here on the special case where $p^+$ lies at $\phi =0$ (and thus $t(p^+) = \pi \ell/2$), as then both $p^+$ and our defect are invariant under the symmetry $\phi \rightarrow -\phi$; see figure \ref{fig:GmDPmD}.  However, recall that (without loss of generality) we have taken $p^+$ to lie to the left of $i^0$ as shown in the figures; i.e., $p^+$ lies at smaller values of $\phi$ and thus of the original global coordinate $\bar \phi = \phi/\alpha$.  But we have $\bar \phi(p_+) =0$ and $\bar \phi(i^0) = \frac{\pi}{2\alpha}$.  Consistency then imposes $\bar \phi(i^0) < \pi$, or $\alpha < 1/2$, which requires $\Gamma < \pi$.  Attempting to allow more general $\Gamma$ while keeping fixed $\phi(p^+) =0$ would allow $p^+$ and $i^0$ to be connected by timelike curves.  Computations for general $p^+$ which allow $\Gamma \in (0,2\pi)$ are relegated to appendix \ref{sec:Append_GeneralHorizons}, though the result will be reported as \eqref{eq:general}.

Our choice $\phi(p^+)=0$ requires $\tau(p)=x(p)$, so setting $z=0$ in \eqref{eq:tant}, \eqref{eq:tanphi} yields
\begin{equation}
\label{eq:tau}
\tan \frac{t}{\ell }=\frac{2\tau }{\ell }=- \cot \phi .
\end{equation}
Since the future horizon $H^+_{bulk}(p)$ is the boundary of  $I^-_{bulk}(p^+)$, its description in the GmD frame will then enjoy this symmetry as well.

\begin{figure}[t]
\centerline{
\includegraphics[width=0.75\textwidth]{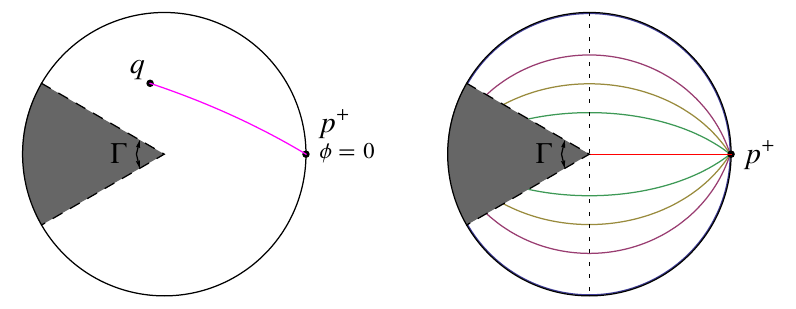}
}
\caption{{\bf Left:}  A projection of the GmD description onto the $(r,\phi)$ surface. Shown are $p^+$ and a general bulk point $q$ in its causal past having $\phi(q) \ge 0$. The causal curve linking them (magenta) may be chosen to avoid the defect (dashed lines).   {\bf Right:} The same projection showing generators of $H^+_{bulk}$; i.e., of the past light cone from $p^+$ with $t(p^+) = \pi \ell/2$, $\phi(p^+)=0$. In empty AdS$_3$, all generators would reach the vertical (dotted) reference line at the same global time $t=0$.  So for $\Gamma < \pi$, when traced backward from $p^+$ the caustic forms at GmD time $t=0$ when the the central $\phi=0$ generator (red) reaches the defect.  Decreasing $t$ further, the caustic moves out from the center.  It finally hits the conformal boundary at $t = - \pi/2 + \Gamma/2$ when the point $p$ reaches the defect.  In contrast, for $\Gamma \ge \pi$ (not shown), tracing the generators backwards from $p^+$ one finds the caustic to form first at the boundary as measured by the GmD time $t$.
 }
\label{fig:GmD}
\end{figure}

To begin, consider any point $q$ in $I^-_{bulk}(p^+)$. Using \eqref{eq:AdS3}, we may think of $q$ as a point in empty AdS$_3$; see figure \ref{fig:GmD} (left).  Without loss of generality we take $\phi(q) \ge 0$, so that also $\phi(q) \le \pi - \Gamma/2$.   Such $q$ can in fact be connected to $p^+$ by a causal curve that satisfies $0 \le \phi \le  \pi - \Gamma/2$ everywhere.  This curve does not pass through the defect and so defines a causal curve in empty global AdS$_3$.    Similarly, any causal curve connecting $q$ and $p$ in global AdS$_3$ can be deformed so as to avoid the defect.  So $H^+_{bulk}(p)$ in the conical defect spacetime is precisely the restriction to $\phi \in [-\pi + \Gamma/2, \pi- \Gamma/2]$ of the empty AdS$_3$ Rindler horizon.  In particular, the caustic on $H^+_{bulk}(p)$ is precisely the intersection of \eqref{eq:globHor} with our defect.  For later use, we
note that with our conventions the global coordinates of points on this horizon satisfy
\begin{equation}
\label{eq:globHor}
 r=\frac{\ell \gamma}{\cos \frac{t}{\ell }},
\ \
 \sin \phi = \frac{\sinh \frac{\eta}{\ell}  \cos \frac{t}{\ell }}{\gamma},  \ \  \cos \phi =\frac{\cosh \frac{\eta}{\ell}  \sin \frac{t}{\ell } }{\gamma},
\end{equation}
where $\eta$ is a parameter labelling generators and in terms of which the induced metric on $H^+_{bulk}(p)$ is $ds^2 = d\eta^2$.  We have also introduced the quantity
$\gamma = \sqrt{\sinh^2 \frac{\eta}{\ell} + \sin ^2\frac{t}{\ell }}$; see appendix \ref{AdS3rev} for further explanation.

To understand $dL_c/dt$, consider tracing the horizon generators backward from $p^+$.
It is useful to note that,
due to the $\phi \to -\phi$ symmetry of our setting, one of the generators lies at $\phi=0$.
Since $\Gamma < \pi$, familiar causal properties of empty AdS$_3$ then imply it to be the first to reach the defect as measured by the GmD time $t$; see figure \ref{fig:GmD} (right).  Having placed $p^+$ at $t=\pi \ell/2$, this occurs at $t=0$.  Since generators reach the defect at the same rate on both sides, for $t < 0$ we have $dL_c/dt =2 d\eta/dt$ in terms of the rate at which the caustic travels a proper distance $\eta$ along the AdS$_3$ Rindler horizon.  Computing $d\eta/dt$ is then simply a matter of setting $\phi = - \pi + \Gamma/2$, noting that the last two equations in \eqref{eq:globHor} give
\begin{equation}
\label{HorIntDef}
\tanh \frac{\eta}{\ell} = \tan \frac{\Gamma}{2} \tan \frac{t}{\ell},
\end{equation}
and differentiating to obtain
\begin{equation}
\label{detadt}
\frac{d\eta}{dt} = \frac{\sin \Gamma}{\cos \frac{2t}{\ell} + \cos \Gamma}.
\end{equation}

To find the remaining factor $dt/d\lambda$, we must locate the intersection of the above horizon (the past light cone of $p^+$) with the corresponding future light cone of $p^-$ and the defect.  Again, the calculation is facilitated by the fact that we wish to compute $Q$ as defined in the PmD frame.  This means that $\lambda$ can be taken to be the PmD coordinate $\tau$ of $p$, and that $p_-$ is the point on the right past null infinity of the PmD Poincar\'e patch (i.e., with $x = -\tau = \infty$) that is null related to $p$; see figure \ref{fig:GmD}.  As shown in the figure \ref{fig:GmDPmD}, we wish to allow $p$ to range over a null line that runs from $p^+$ to the defect.  Since both $p$ and $p^-$ remain away from the defect,  we can obtain $p^-$ from $p$ using the GmD null translation $(t, \phi) \rightarrow (t - \pi \ell/2, \phi + \pi/2)$, which also maps $p^+$ to $i^0$. Note that acting with this translation on \eqref{eq:tau} gives the GmD coordinates $(t_-,\phi_-)$ of $p_-$ in terms of the affine parameter $\lambda$ defined by $\tau(p)$:
\begin{equation}
\label{eq:tauminus}
- \cot \frac{t_-}{\ell }=\frac{2\lambda }{\ell }= \tan \phi_- .
\end{equation}

\begin{figure}[t]
\centerline{
\includegraphics[width=0.75\textwidth]{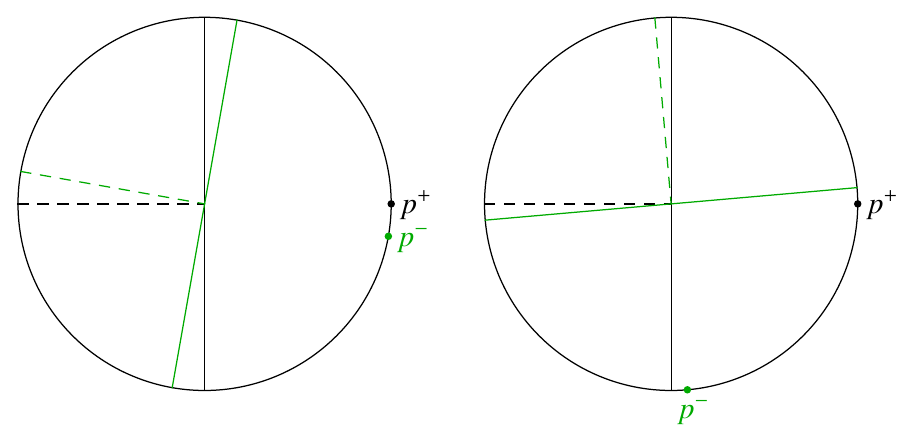}
}
\caption{Since we are interested only in $\phi _-\in (-\pi /2,0]$, even with the largest interesting defect angle ($\Gamma = \pi$), an extra defect (green) centered opposite $p^-$ does not affect computations with $\phi \le 0$ so long as we also work outside the original defect (black) opposite $p^+$.}
\label{fig:twodefects}
\end{figure}

Furthermore,  in empty AdS$_3$ the future light cone of $p^-$ can be found by acting on the time-reverse of \eqref{eq:globHor} with the $t$ and $\phi$ translations that move the time-reverse of $p^+$ to $p^-$. These are respectively the operations $t \rightarrow t + \pi \ell/2 + t_-$ and $\phi \rightarrow \phi + \phi_-$ where $(t_-,\phi_-)$ are the global coordinates of $p_-$.  As usual, this reasoning remains valid in our GmD coordinates so long as there is no interference from the defect.  In this step, there is an additional subtlety that the above translations effectively move the defect so that it becomes centered on line $\phi = \phi_-+ \pi$ opposite $p^-$.  On the other hand, we also wish to obtain relations valid in our GmD frame where the defect is fixed to be centered on $\phi = \pi$ so that it lies opposite $p^+$.  But by construction, we have $\phi_- >-\pi/2$. And as noted above, the caustic exists only for $t(p) \le 0$ (i.e., for $\lambda \le 0$) and thus for $\phi_- \le 0$; see again figure \ref{fig:GmD} (right).  Figure \ref{fig:twodefects} then makes clear that an extra defect located opposite $\phi_-$ has no effect so long as long we describe the caustic as lying at negative $\phi $ (and thus as $\phi = -\pi + \Gamma /2$). Some algebra then locates the intersection of the light cones from $p^\pm$ (i.e., the CHI surface) at
\begin{equation}
\cot \frac{t}{\ell} = \frac{\ell}{\lambda} + \tan \phi .
\end{equation}
Setting $\phi = - \pi + \Gamma/2$ and differentiating yields
\begin{equation}
\label{dtdl}
\frac{dt}{d\lambda} = \frac{\ell^2}{\ell^2 + \lambda^2 +2 \ell \lambda \tan \frac{\Gamma}{2} + \lambda^2 \tan^2 \frac{\Gamma}{2}}.
\end{equation}
Combining \eqref{detadt} and \eqref{dtdl} using standard trigonometric identities we finally obtain
\begin{equation}
\label{eq:areachange}
\frac{dL_c}{d\lambda}= 2\frac{d\eta}{dt}\frac{dt}{d\lambda}=\frac{1}{\frac{\lambda }{\ell }+\frac{1}{2} \cot \frac{\Gamma }{2}},\text{ for }\lambda <0.
\end{equation}

Since $\frac{dL_c}{d\lambda }$ vanishes for $\lambda >0$, a short computation shows our result to satisfy
\begin{equation}
\label{eq:delta}
\frac{d^2L_c}{d\lambda ^2}+\frac{1}{\ell }\left(\frac{dL_c}{d\lambda }\right)^2=-2\tan \frac{\Gamma}{2}\delta \left(\lambda \right),
\end{equation}
where the right-hand-side contains a Dirac delta function.  We thus see that the coarse-grained QNEC \eqref{2dQNEC} holds everywhere.

Although we have focused above on the case where $\phi_+ := \phi(p^+)=0$ (and thus $t_+ : = t(p^+) = \pi\ell/2$),  other choices for $p^+$ can be dealt with similarly.  One need only
apply appropriate $t,\phi$ translations to \eqref{eq:globHor} and to take appropriate are with the defect.  Some key steps are outlined in appendix \ref{sec:Append_GeneralHorizons}.

To explain the main results from that appendix, we mention that taking $\phi_+ \neq 0$ allows considerations of general $\Gamma < 2\pi$.  The constraint that no timelike curve can connect $i^0$ to $p^+$ turns out to require only $-\frac{\pi }{2}+\frac{\Gamma }{2}<\phi _+<\frac{\pi }{2}$, allowing the original global coordinate $\bar \phi_+$ of $p^+$ to range over the full desired interval from $\bar \phi(i^0) - \pi$ to  $\bar \phi(i^0)$ for all $\Gamma < 2\pi$.

When in addition $\phi _+<\frac{\pi }{2}-\frac{\Gamma }{2}$, appendix \ref{sec:Append_GeneralHorizons} finds

\begin{equation}
\label{eq:general}
\frac{d^2L_c}{d\lambda ^2}+\frac{1}{\ell }\left(\frac{dL_c}{d\lambda }\right)^2=-\frac{2}{\cot \frac{\Gamma}{2}-\tan \phi _+}\delta \left(\lambda +\ell \tan \phi _+\right).
\end{equation}
Here $\lambda$ is again the PmD affine parameter associated with $\tau(p)$. As in \eqref{eq:delta}, the coefficient of the Dirac delta function on the right-hand-side is negative so that the coarse-grained QNEC holds for all $\lambda$.

For the remaining case $\phi _+ \ge \frac{\pi }{2}-\frac{\Gamma }{2}$, one finds a new behavior where the largest value of the PmD $\lambda$ on the caustic occurs at the conformal boundary instead of in the interior. Since the horizon length diverges near the boundary, this means that $L_c$ (defined by integrating $dL_c/dt$) is constant when there is no caustic and has an infinite discontinuity when the caustic forms.  This case is too singular to allow a definitive analysis of \eqref{2dQNEC}, though \eqref{2dQNEC} is certainly consistent with simple regularizations of this result.

\subsection{A physical normalization for $Q$}
\label{sec:Append_GenuinePoincare}

For convenience, the computations of our QNEC quantity $Q$ above were performed in the PmD frame.  This is sufficient to verify \eqref{2dQNEC} in any frame, as $Q$ transforms as a scalar of weight $2$; multiplying by $\Omega^{-2}$ cannot change the sign of $Q$ for any conformal factor $\Omega$.

However, it is interesting to ask what structure might be encoded in the delta-function on the right-hand-side of \eqref{eq:delta} or \eqref{eq:general}.  One might expect the coefficient to be easiest to interpret in a frame where the right-moving null line through $p$ is a Killing horizon.  This was the Poincar\'e frame $(\bar \tau, \bar x)$ discussed at the very beginning of section \ref{sec:setting}.  Denoting $Q$ in this frame by $Q_P$, we have $Q_p = Q_{PmD} \left(\frac{d\lambda}{d\bar \lambda} \right)^2$ in terms of the Poincar\'e frame affine parameter $\bar \lambda$.    Along the original line $\tau = x$ we have $\bar \tau = \bar x - \ell \cot \frac{\pi }{2 \alpha }$,  so the analogue of \eqref{eq:tau} is
\begin{equation}
\tan \left[\bar \phi -\frac{\pi }{2}\left(\frac{1}{\alpha }-1\right)\right]=\frac{-\ell ^2+\left(\bar \lambda +\ell \cot \frac{\pi}{2\alpha }\right)^2-\bar \lambda ^2}{2\ell \left(\bar \lambda +\ell \cot \frac{\pi}{2\alpha }\right)}.
\end{equation}
Combing this with \eqref{eq:tau} for $\lambda = \tau$ and the relation $\bar \phi = \phi/\alpha$ yields
\begin{equation}
\bar \lambda =\frac{\ell}{2}\left[\cot \left(\frac{1}{\alpha }\text{arccot}\frac{2\lambda }{\ell }\right)-\cot \frac{\pi}{2\alpha}\right],  \ \ \
\frac{d\bar \lambda }{d\lambda }\left| _{\lambda =0} \right.=\frac{1}{\alpha {{\sin }^{2}}\frac{\pi }{2\alpha }},
\end{equation}
where we remind the reader that $\alpha = 1 - \frac{\Gamma}{2\pi}$.  Unfortunately, multiplying \eqref{eq:delta} by $\left( \frac{d\bar \lambda }{d\lambda }\left| _{\lambda =0} \right.\right)^2$ does not significantly simplify the result.

\section{Discussion}
\label{sec:discussion}

We have shown conical defect AdS$_3$ spacetimes to define dual states of holographic CFTs that satisfy a quantum null energy condition \eqref{2dQNEC} in terms of the coarse-grained entropy $S_{QFT}^{coarse}$ defined by the causal holographic information (CHI) of \cite{Hubeny:2012wa}.    In particular, the right-hand side is a delta-function with negative coefficient supported at the instant where the CHI surface reaches the conical singularity.  This result is a surprise, as it is determined entirely by the rate at which horizon area\footnote{Actually length, due to the low dimension.} emerges from a caustic. We emphasize that the caustic persists for all $\lambda \le 0$.  So while the delta-function is associated with what one might call the termination of the caustic, the fact that \eqref{eq:delta} vanishes for negative $\lambda$ represents a novel cancellation along the caustic's entire length. We have found no prior results controlling the rate of area emergences.   A general proof of our coarse-grained QNEC would then imply unprecedented constraints on the growth of caustics.  While a CHI QNEC is not be logically required by either the conjecture of \cite{Hubeny:2012wa} that CHI represents a coarse-grained entropy or by the fine-grained QNEC conjecture of \cite{Bousso:2015mna}, it represents an interesting new conjecture that fuses the above two.

\begin{figure}[t]
\centerline{
\includegraphics[width=0.75\textwidth]{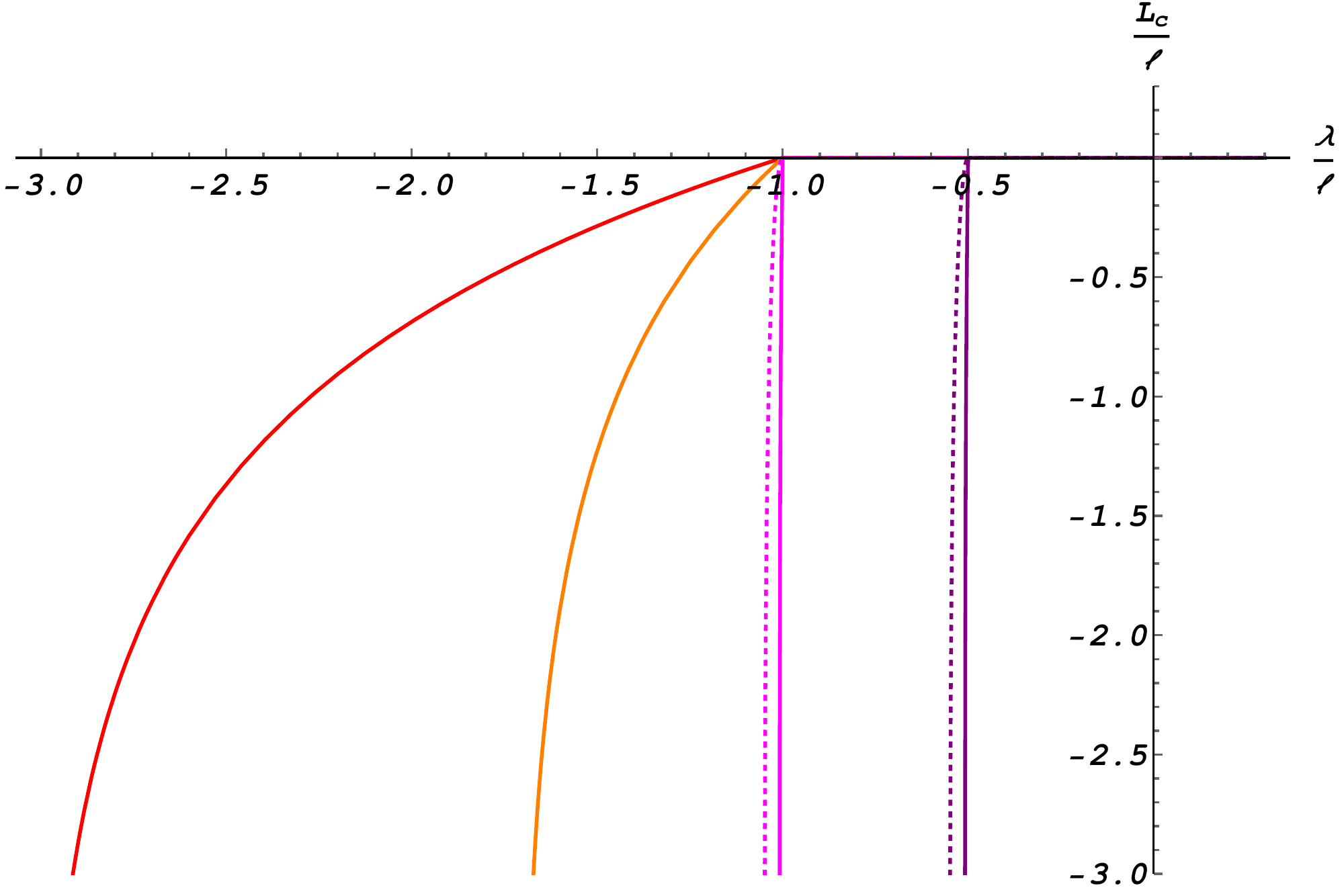}
}
\caption{$L_c$ as a function of $\lambda _{PmD}$ for $\phi _+=\pi /4$. From left to right, $\Gamma = \frac{\pi}{8}, \frac{\pi}{4}, \frac{\pi}{2}, \pi$. The right-most two have $\Gamma \ge \pi- 2\phi_+$ and thus transition suddenly at some $\lambda_c$ to finite values from $L_c=-\infty$.   Regulated curves may then be defined by replacing $L_c$ with e.g. $\frac{1}{\ell }L_{c,reg} = \ln\left[\cos \left(\frac{\pi}{2}\frac{\lambda - \lambda_c}{\ell \epsilon} \right) \right]$ for $\frac{\lambda_c}{\ell } - \epsilon < \frac{\lambda }{\ell } < \frac{\lambda _c}{\ell }$. Such curves approach the solid lines as $\epsilon \rightarrow 0$ and give a regulated $Q_{PmD}$ equal to $\frac{\pi}{32\epsilon^2G_{bulk}}$  (i.e., satisfying \eqref{2dQNEC}) for all $\frac{\lambda_c}{\ell } - \epsilon < \frac{\lambda }{\ell } < \frac{\lambda _c}{\ell }$.  The dashed lines are sample such curves with $\epsilon=0.05$.    Plots for other values of $\phi_+$ are similar.}
\label{fig:SingS}
\end{figure}

Our work concentrated on the case $\Gamma  < \pi -2 \phi _+$ where \eqref{2dQNEC} holds in the usual sense of distributions.  In the complimentary case $\Gamma  \ge \pi -2 \phi _+$, our $S_{QFT}^{coarse}$ transitions suddenly from $- \infty$ to a finite constant.  This behavior is too singular to allow a definitive analysis of \eqref{2dQNEC} at the disconinuity, though simple regularizations of this result certainly satisfy \eqref{2dQNEC}; see figure \ref{fig:SingS}.  Note that this singular behavior of $S_{QFT}^{coarse}$ results from the form of the asymptotic gravitational field and has nothing to do with the conical singularity in the spacetime.  In particular, the singular behavior of $S_{QFT}^{coarse}$ would remain if our conical deficit were replaced by a star of the same mass.  This suggests that the QNEC has meaning (and should be expected to hold) only when $Q$ is finite.

It is interesting that in our simple examples $S_{QFT}^{coarse}$ in fact saturates \eqref{2dQNEC} except at the special point where the CHI surface reaches the conical singularity. For completeness, we mention that the behavior of the fine-grained HRT entropy is similar.  Indeed, it is known that HRT surfaces in conical defect spacetimes do not reach the conical singularity; see e.g. \cite{Balasubramanian:2014sra}  for a discussion of the time-symmetric (RT) case.  As a result, any HRT surface is an extremal surface in empty AdS$_3$. Away from HRT phase transitions where the extremal surface changes discontinuously, derivatives of the HRT entropy $S_{HRT}$ will thus also agree with derivatives in empty AdS$_3$.  But in empty AdS$_3$ we may compute \eqref{2dQNEC} in the Poincar\'e conformal frame where all contributions to \eqref{2dQNEC} vanish explicitly.  As a result, in our conical defect spacetimes \eqref{2dQNEC} yields a delta-function governed by the discontinuity in $dS_{HRT}/d\lambda$ at the phase transition.  Defining $L = 4G_{bulk} S_{HRT}$ and taking $p$ on the null line $\tau =x$, a calculation yields
\begin{equation}
\label{finedelta}
\frac{d^2L}{d\lambda ^2}+\frac{1}{\ell}\left(\frac{dL}{d\lambda }\right)^2=-\sin \Gamma \delta \left(\lambda -\frac{\ell}{2}\tan \frac{\Gamma}{2}\right),
\end{equation}
where again $\lambda =\tau (p)$. Since the support of \eqref{finedelta} is always later in time than that of \eqref{eq:delta},  at all $\lambda$ the coarse-grained entropy equals or exceeds the fine-grained entropy as required by \cite{Hubeny:2012wa,Maximin}.  Unfortunately, even in the naturally preferred conformal frame studied in section \ref{sec:Append_GenuinePoincare}, we find no clean interpretation of the coefficient on the right-hand-side for either \eqref{finedelta} or our coarse-grained result \eqref{eq:delta}.

It is interesting to ask just how generally one may expect the QNEC to hold in either our coarse- or the original fine-grained formulation.  Based on its relation to possible covariant entropy bounds, the fine-grained QNEC was conjectured in \cite{Bousso:2015mna} to hold in all backgrounds.  On the other hand, the relation to the GSL \cite{Wall:2011kb,coarse-grainedGSL} is direct only on Killing horizons.  For $d=2$ CFTs, any point may be taken to lie on a Killing horizon by making a suitable change of conformal frame.  But this is not the case for non-conformal $d=2$ theories or in higher dimensions.  As a result, both forms of the QNEC should be explored in these more general contexts.  One should bare in mind that our renormalization of the coarse-grained CHI entropy using HRT counterterms is only known to be valid on Killing horizons in settings that approach equilibrium (see \cite{coarse-grainedGSL} and appendix \ref{sec:Append_2dGSL}), and indeed that it fails more generally \cite{CHI}, though it may still be interesting to ask if the general non-local divergences satisfy a QNEC.

Returning to the special cases studied here, recall that the analysis of our coarse-grained QNEC relied only on the behavior at caustics; there was no contribution from the bulk expansion along individual generators of our horizon.  It will clearly be useful to study further examples where this bulk expansion plays a role.   And since three-dimensional gravity has many special properties, higher dimensional spacetimes may be quite different.  But while much can be learned from computations in various special cases, a general understanding of the properties of CHI may require development of new tools that better describe the rate at which horizons grow along caustics.    If such tools could establish the general validity of our coarse-grained QNEC (or perhaps just on Killing horizons and in situations approaching equilibrium), they would provide striking further evidence that CHI defines a useful notion of coarse-grained holographic entropy.

\section*{Acknowledgements}
ZF is pleased to thank Han Liu for helpful discussions. DM thanks Aron Wall for many related conversations. This work was supported in part by the Simons Foundation and by funds from the University of California.

\appendix

\section{The Coarse-grained GSL for holographic d=2 QFTs}
\label{sec:Append_2dGSL}

While Einstein gravity becomes trivial for $d=2$, one can still formulate a GSL in dilaton gravity theories, in which the dilaton field plays the role of the area \cite{Fiola:1994ir}. We now derive a corresponding $d=2$ holographic coarse-grained GSL at leading non-trivial order in $G_2$, where $G_2$ is the coupling to dilaton gravity.  Note that this $G_2$  has nothing to do with the parameter $G_{bulk}$ of the bulk holographic dual.  Indeed, $G_2$ will make no further appearance beyond the next two paragraphs below\footnote{The material in this appendix was originally prepared for a paper to be co-authored with Aron Wall. We thank him for his permission to use it here.}.

Our basic setting is similar to that of \cite{coarse-grainedGSL}.  In particular, we study a unitary 1+1 holographic QFT on a Killing horizon $H^{(0)}$ of a general curved spacetime with metric $g^{(0)}_{\alpha \beta}$.  This $H^{(0)}$ then defines a corresponding bulk event horizon $H$; see section 4 of \cite{coarse-grainedGSL} for details.   We also assume the system to reach equilibrium in the far future.  However, while the only tensor source we allow is the metric, we will allow scalar sources of non-negative dimension below.  We will also take care to allow general bulk bosonic matter fields so that our version of the result holds beyond the so-called universal sector of holographic theories. Due to our classical treatment of the bulk, we set all bulk fermions to zero.

As explained in the introduction, we proceed by writing
\begin{equation}
\label{split}
\frac{d}{d\lambda} S^{coarse}_{QFT} = \frac{1}{4G_{bulk}} \left( \frac{d}{d\lambda} A_{\text{Hawking}} + F \right),
\end{equation}
where $\frac{d}{d\lambda} A_{\text{Hawking}}$ is a term guaranteed to be non-negative by the Hawking area increase law in the bulk and $F$ is the flux through the boundary of bulk horizon area (here a length since $d=2$).  We will show below that $F$ for which $\frac{F}{4 G_{bulk}}$ saturates \eqref{2dQNEC}; i.e. for which
\begin{equation}
\label{2dQNECadiabatic}
8 \pi G_{bulk} T_{\alpha \beta} k^\alpha k^\beta - \left(  \frac{d}{d\lambda}  F + \ell^{-1} F^2 \right) = 0
\end{equation}
in terms of the bulk AdS length scale $\ell$.   Combing this with \eqref{split} and the dilaton equivalent of the linearized Raychaudhuri equation (see e.g. \cite{Wall:2011kb}) then yields the leading order $G_2 \rightarrow 0$ GSL.  These steps are described for dilaton in \cite{Wall:2011kb} and are in direct parallel to that used for higher dimensions in \cite{coarse-grainedGSL}.  They also correspond to standard derivations of the physical process version of the first law of black hole mechanics \cite{Carter,Wald:1995yp,Jacobson:1999mi,Amsel:2007mh}.  As a result, we will not repeat them here. Instead, we focus below on establishing \eqref{2dQNECadiabatic}.

\subsection{Divergences}
\label{appdiv}
A simplifying feature of $d=2$ is that there are tight constraints on the possible divergences. We will be interested in divergences in the stress tensor, but these are determined by divergences in the action (or, more generally for non-Lagrangian theories,  in the partition function).  Recall that by definition any such action divergence will have total dimension $d=2$  given by the sum of the dimensions of any operators, derivatives, and divergent coefficients.  The operator dimension is strictly positive in holographic theories\footnote{The general CFT unitarity bounds allow scalar operators of dimension $\frac{d-2}{2}=0$, but they also require these operators to be free fields.  On the other hand, free fields should not be holographic, and indeed \cite{Andrade:2011dg} showed directly that holographic scalars of dimension $\frac{d-2}{2}$ violate unitarity.}, so the only allowed action divergence involving the Riemann tensor is a logarithmic divergence proportional to the integral of the Ricci scalar $R^{(0)}$ of $g^{(0)}_{\alpha \beta}$.  This is a topological invariant and cannot contribute to $T_{\alpha \beta}$.

Covariance requires action divergence terms not involving the Riemann tensor to be of the form
\begin{equation}
\label{Oterm}
\int_B \sqrt{g^{(0)}} {\cal O},
\end{equation}
where ${\cal O}$ is a scalar operator of dimension $\Delta_{\cal O} \le 2$.  Since stress tensor divergences are variations of such terms with respect to $g^{(0)}_{\alpha \beta}$, one should ask if this ${\cal O}$ might implicitly depend on $g^{(0)}_{\alpha \beta}$.  Due to covariance, this can be the case only if ${\cal O}$ is constructed from bosonic (and thus tensor) bulk fields using the metric and covariant derivatives\footnote{In the presence of non-scalar operators, one may ask what should be held fixed in this variation.  In fact \cite{Hollands:2005ya} one should introduce frame fields and fix all tensor components with internal (tangent-space) indices.}.  Since $d=2$, anti-symmetric parts of tensors are dual to scalars and we need only consider symmetric tensors.  But the unitarity bounds \cite{di2009conformal} forbid symmetric tensors of rank $2$ from having dimension less than $2$, and $\Delta =2$ is allowed only for traceless tensors of rank $2$.  Covariance and their vanishing trace then prohibit such operators from appearing in our term.  Note that covariant derivatives of scalars ($\nabla_a \phi = \partial_a \phi$) do not in fact depend on the metric, so again using covariance the only possibly non-trivial case is ${\cal O} = \Phi \nabla_a j_a$ for some scalar source $\Phi$.  But we may then integrate \eqref{Oterm} by parts to move the derivative onto $\Phi$ so that it becomes independent of the metric.  Again, only the explicit $\sqrt{-g^{(0)}}$ in \eqref{Oterm} contributes when we take variations with respect to $g^{(0)}_{\alpha \beta}$.

It follows that all divergent counterterms $T^{CT}_{\alpha \beta}$ in the stress tensor are proportional to $g^{(0)}_{\alpha \beta}$.  As a result, $T^{CT}_{\alpha \beta}k^\alpha k^\beta =0$ for any null vector $k^\alpha$ of $g^{(0)}_{\alpha \beta}$ and divergent counterterms cannot contribute to \eqref{2dQNEC}.

We should also enumerate possible divergent counterterms in the HRT entropy.  As stressed in \cite{Taylor:2016aoi}, the Maldacena-Lewkowycz argument \cite{Lewkowycz:2013nqa} suggests that these are again determined by counterterms in the action, and indeed the logarithmic Ricci-scalar counterterm gives the standard logarithmic counter term shown in \eqref{2dCHI}.   We simply mention that no further divergences can arise from moving beyond the universal sector as the leading term in the Fefferman-Graham expansion cannot be affected by bulk matter fields of positive dimension and no other terms can give divergences in the HRT entropy.  This is equivalent to the observation of \cite{Marolf:2016dob} that for $d=2$ state-dependent divergences in the entropy can arise only in the presence of operators with strictly vanishing conformal dimension.  We note that on any surface of constant Fefferman-Graham $z$ coordinate the counterterm in \eqref{2dCHI} is independent of position and may be ignored when computing derivatives.  As a result, divergences cannot contribute to changes in $S^{coarse}_{QFT}$.

\subsection{Main Argument}

We are now nearly ready to derive the critical relations \eqref{split} and  \eqref{2dQNECadiabatic}.  We work with the gravitational dual of our 1+1 holographic CFT, and we take this dual to be locally asymptotic to AdS$_3$.  In particular, we impose Fefferman-Graham gauge near the boundary so that the bulk metric takes the form
\begin{equation}
\label{GFG}
G_{AB} dX^A dX^B = \frac{\ell^2}{z^2} \[ dz^2 + g_{\alpha \beta}(z)  dx^\alpha dx^\beta \],
\end{equation}
with $g_{\alpha \beta}(z) \to g^{(0)}_{\alpha \beta}(z)$ as $z \to 0$.  Spacetimes that are
asymptotic to AdS$_3 \times X$ with $X$ compact can also be treated using Kaluza-Klein reduction, though we will not comment further on this case.

As noted above, so long as we regulate our calculation by working inside a surface of constant Fefferman-Graham coordinate $z=z_0$, divergent counterterms cannot contribute to  changes in $S^{coarse}_{QFT}$. So it suffices to study $S^{coarse}_{QFT}$ using only the bare term in \eqref{2dCHI} (given by the area of our causal horizon with $z > z_0$) and the bare stress tensor.
We find it useful to follow \cite{Balasubramanian:1999re} in defining the bare stress tensor by varying the Einstein-Hilbert action with Gibbons-Hawking term for the region $z > z_0$.  This gives
\begin{equation}
\label{bareT}
T^{bare}_{\alpha \beta} (z_0) = \frac{\ell^{d-2}}{8 \pi G_{d+1} z^{d-2}} \left[K_{\alpha \beta} - \frac{\ell^2}{z^{2}} K g_{\alpha \beta}(z_0)\right],
\end{equation}
where  $K_{\alpha \beta}$ is the extrinsic curvature of the surface $z=z_0$ as defined by the inward-pointing normal, and $K = \frac{z^2}{\ell^2} g^{\alpha \beta}(z_0) K_{\alpha \beta}$ is its trace with respect to $G_{AB}$.  Note that the arguments of section \ref{appdiv} and equations \eqref{GFG}, \eqref{bareT} imply that we may write
\begin{equation}
\label{bare}
T^{bare}_{\alpha \beta}(z_0) = T^{finite}_{\alpha \beta}(z_0) +  \frac{f(z_0)}{8\pi G_3z^{2}} g^{(0)}_{\alpha \beta},
\end{equation}
where as $z_0 \to 0$ we have both $f(z_0) \to 1$  and $T^{finite}_{\alpha \beta}(z_0) \rightarrow T_{\alpha \beta}$, where $T_{\alpha \beta}$  is the (manifestly finite) renormalized boundary stress tensor.

We now turn to the (regulated) bare entropy $S_{bare} = \frac{1}{4G_{3}} \text{Length}(C_{z>z_0})$, where $C$ is a cut of the bulk horizon ending at the desired boundary points and we include only the length in the region $z>z_0$ inside our regulating surface. The rate of change of this area will enter into $\frac{d}{d\lambda}S_{QFT}$.  As in \cite{coarse-grainedGSL}, $\frac{d S_{bare}}{d\lambda}$ may be divided into two contributions.  The first is the rate at which area is created in the bulk as determined by both the local divergence of tangents to the generators of $H$ and the rate at which generators are added to the horizon.  Since we assume the QFT to reach equilibrium in the far future, the bulk must settle down to a stationary black hole.  That the bulk area creation term is non-negative is then just the usual Hawking area theorem \cite{Hawking:1971tu}.  After multiplying by $\frac{1}{4G_{3}}$, we call this $\frac{d}{d\lambda}S_{QFT,\ non-dec}$ as explained in the introduction.

The remaining contribution $4 G_{3} \frac{d}{d\lambda} S_{QFT, \ adiabatic}$ to $\frac{d}{d\lambda} {\rm Length}(C_{z>z_0})$ is then the rate at which length flows inward through the cutoff surface $z=z_0$ as measured by an affine parameter $\lambda$ of $g^{(0)}_{\alpha \beta}$. We will call this flux of length $F$.  Now, for small $z_0$ the curve $C$ will intersect $z=z_0$ at precisely two points.  For simplicity, let us assume that one of these points remains fixed and that only the other depends on $\lambda$.  There is then only a single point at which length can enter the bulk so we may write
\begin{equation}
\label{FluxC}
F:= \lim_{z_0 \rightarrow 0} U^A n_A,
\end{equation}
where the one-form $n_AdX^A = \frac{\ell}{z}dz$ is the (inward pointing) unit normal to the cutoff surface $z=z_0$ and $U^A$ are the null tangents to the generators of $H$ that satisfy $U^A\partial_A : = \frac{d}{d\lambda}$.  Here we have used the fact that -- since we hold fixed one endpoint of $C$ -- the CHI prescription defines a unique cut $C$ of the bulk horizon for each point $p$ at which the other endpoint may lie in the boundary horizon $H^{(0)}$.  This fact allows us to extend the definition of the $\lambda$ from $H^{(0)}$ to the full bulk horizon $H$. Though while $\lambda$ is an affine parameter on the generators of $H^{(0)}$, it is not necessarily affine on bulk generators of $H$.

Now, the  timelike nature of the boundary and the results of Gao and Wald \cite{Gao:2000ga} imply that the actual bulk horizon $H$ approaches $H^{(0)}$ as $z \rightarrow 0$.  So up to normalization we have $U^\alpha\partial_\alpha \rightarrow k^\alpha \partial_\alpha$ as $z \to 0$.
As a result, it is convenient to label points on such generators using an affine parameter $\tilde \lambda$ defined by the rescaled bulk metric $\tilde G_{AB}$ defined by
\begin{equation}
\label{tildeG}
\tilde G_{AB} dX^A dX^B = \frac{z^2}{\ell^2} G_{AB} dX^A dX^B = dx^2 + g_{\alpha \beta}(z) dx^\alpha dx^\beta,
\end{equation}
whose causal structure agrees with that of $G_{AB}$ but for which the counting of powers of $z$ is somewhat more direct. We normalize $\tilde \lambda$ so that $\tilde \lambda \rightarrow \lambda$ as $z \rightarrow 0$.  The associated null tangents are $\tilde U^A = \frac{d\lambda}{d\tilde \lambda}U^A$, so since $\frac{d\lambda}{d\tilde \lambda} \to 1$ we may rewrite \eqref{FluxC} as
\begin{equation}
\label{FluxC2}
F  = \lim_{z_0 \rightarrow 0} \tilde U^A n_A.
\end{equation}

We now study the evolution of $\tilde U^A \tilde n_A$ with $\tilde \lambda$. It will be useful to express the rate of change $\frac{d}{d \tilde \lambda} \tilde U^A \tilde n_A$ in terms of the extrinsic curvature $\tilde K_{AB}$ of the cutoff surface $z=z_0$ as defined by the rescaled metric $\tilde G_{AB}$.  Here we use the conventions of \cite{Wald:1984rg}, taking $\tilde K_{AB}$ to be a degenerate tensor in the full spacetime.
 In general, $\tilde K_{AB}$ is given by projecting $\tilde \nabla_A \tilde n_B$ into the surface on the first index, where $\tilde \nabla_A$ and  $\tilde n_B$ are the bulk covariant derivative and unit normal defined by $\tilde G_{AB}$.  But our coordinates $X^A = (z,x^\alpha)$ are Gaussian normal with respect to $\tilde G_{AB}$, so $\tilde n^A$ satisfies the geodesic equation $\tilde n^B \tilde \nabla_B \tilde n^A =0$ and the projection is trivial.  We may thus write simply $\tilde K_{AB} = \tilde \nabla_A \tilde n_B$.
The geodesic equation for $\tilde U^A$ then gives
\begin{equation}
\label{ddl}
\frac{d}{d \tilde \lambda} \left( \tilde U^A \tilde n_A\right) : = \tilde U^B \tilde \nabla_B \left( \tilde U^A \tilde n_A \right) =  \tilde U^A \tilde U^B \tilde \nabla_B  \tilde n_A  =  \tilde U^A \tilde U^B \tilde K_{AB}.
\end{equation}
And since the extrinsic curvature $K_{AB} = \frac{1}{2} \pounds_{n} (G_{AB} - n_A n_B)$ defined by the physical bulk metric $G_{AB}$ satisfies $\tilde K_{AB} = \frac{z}{\ell^2}\left[ (G_{AB} - n_A n_B) +  \ell K_{AB}\right]$,
we may use \eqref{bareT} and the fact that $\tilde U^A$ is null with respect to $G_{AB}$ to rewrite \eqref{ddl} in the form
\begin{equation}
\label{tddT}
\frac{d}{d \tilde \lambda} \left( \tilde U^A \tilde n_A \right)   = \tilde U^\alpha \tilde U^\beta \frac{8\pi G_{3} z}{\ell} T^{bare}_{\alpha \beta}
- \frac{K\ell+1}{z}   (\tilde U^A \tilde n_A)^2,
\end{equation}
where $K =  G^{AB} K_{AB} \rightarrow  -d/\ell = -2/\ell$ as $z \rightarrow 0$.  Since $\tilde n_A = \frac{z}{\ell} n_A$   and $\tilde n_A \tilde U^A = \tilde U^z : = \frac{dz}{d\tilde \lambda}$, using \eqref{bare} yields
\begin{equation}
\label{ddT2}
\frac{d}{d \tilde \lambda} \left( \tilde U^A n_A \right)   =  8\pi G_{3} \tilde U^\alpha \tilde U^\beta   T^{finite}_{\alpha \beta}
- (K+(2+f(z))/\ell)   (\tilde U^A n_A)^2.
\end{equation}

Equation \eqref{ddT2} is essentially the desired result.  However, before passing to \eqref{2dQNECadiabatic}, we use \eqref{ddT2} to show that $(\tilde U^A n_A)$ is bounded near $z=0$.  Suppose instead that $(\tilde U^A n_A)$ becomes large as $z \to 0$.  Then to good approximation we can ignore the manifestly bounded source term $ 8\pi G_{3} \tilde U^\alpha \tilde U^\beta   T^{finite}_{\alpha \beta}$.  Furthermore, taking $z$ near $0$ allows us to write $f(z) = 1 + \dots $ and $K = -2/\ell +\dots$, where in both cases $\dots$ indicates terms that are subleading by a power of $z$. The remaining homogeneous equation  resembles the source-free Raychaudhuri equation, and in the same way implies that, if $(\tilde U^A n_A) \neq 0$ at some $\lambda$, we must in fact find $(\tilde U^A n_A) \to \pm \infty$ within some $\Delta \lambda$ of order $1/(\tilde U^A n_A) \neq 0$. But since our spacetime is smooth for $z> 0$, the affinely parameterized tangent (and thus also the quantity $(\tilde U^A n_A)$ can diverge only at the boundary. So this then requires our nearby null geodesics to reach the boundary within a small range of $\tilde \lambda$.

Now, as noted above the results of Gao and Wald \cite{Gao:2000ga} forbid our geodesics from reaching the boundary to the future of $H^{(0)}$.  And, since $H^{(0)}$ is a generator of $H$, in the case where it reaches the boundary on $H^{(0)}$ itself we see that our geodesic reaches a caustic.  We may then extend the conformally compactified spacetime to conclude that our geodesic enters the interior of the past of  $H^{(0)}$. Since the region near any point on the boundary is topologically trivial, this means that our geodesic can be smoothly deformed to some timelike curve.   The same conclusion holds (without needing the conformal extension) for any null geodesic reaching the boundary to the past of $H^{(0)}$.
In either case, this requires our geodesic to reach a conjugate point \cite{Hawking:1973uf} within the small region being studied.  And since this occurs for any geodesic near the boundary, there must be a sequence of conjugate points that converge to a point $c$ on $H^{(0)}$.  Continuity of Jacobi deviation vector fields then requires that $c$ be a conjugate point for $H^{(0)}$.  But this is impossible, since $H^{(0)}$ is a future Killing horizon of the boundary spacetime.

It follows that $\tilde U^An_A$ must remain bounded as $z \to 0$ as claimed above.  Under the additional assumption that $\tilde U^An_A$ admits a power series expansion in $z$, it follows that the limit $F = \lim_{z \to 0} U^An_A$ converges, and it makes sense to take $z \to 0$ in \eqref{ddT2}.  Recalling that $K + (2 -f(z))/\ell \rightarrow 1$ and $T^{finite}_{\alpha \beta} \rightarrow T_{\alpha \beta}$ then yields \eqref{2dQNECadiabatic} as desired.

\section{Coordinates and horizons in empty AdS$_3$}
\label{AdS3rev}

We begin by briefly reviewing Rindler horizons in empty AdS$_3$.   Recall that AdS$_3$ may be defined as (the universal cover of) the hyperboloid
\begin{equation}
\label{AdShyp}
(T^1)^2+(T^2)^2-(X^1)^2-(X^2)^2=\ell ^2
\end{equation}
in a 2+2 dimensional Minkowski-like spacetime with metric
\begin{equation}
ds^2_{2+2} = - (dT^1)^2 -(d T^2)^2+ (dX^1)^2+(dX^2)^2.
\end{equation}

In the main text we make use of two standard coordinate systems on AdS$_3$. The first is given by the global coordinates $(t,r,\phi)$ obtained by writing
\begin{align}
\label{eq:globalcoords1}
T_1=\sqrt{\ell ^2+r^2}\cos \frac{t}{\ell },\  &  \ T_2=\sqrt{\ell ^2+r^2}\sin \frac{t}{\ell },
\\
\label{eq:globalcoords2}
X_1=r\cos \phi,\  &  \ X_2=r\sin \phi,
\end{align}
so that the induced metric becomes \eqref{eq:AdS3}.
Pulling out a conformal factor $\frac{\bar r^2}{\ell^2}$ from \eqref{eq:AdS3}, such metrics are naturally associated with a conformal frame in which the boundary metric is $ S^1 \times {\mathbb R}$ where the $S^1$ has radius $\ell$.

The second is given by the Poincar\'e coordinates $(\tau,x,z)$ defined by
\begin{align}
T_1=\frac{z}{2}\left(1+\frac{\ell ^2+x^2-\tau ^2}{z^2}\right),\  &  \ T_2=\frac{\ell \tau }{z},
\\X_1=\frac{\ell x}{z},\  &  \ X_2=\frac{z}{2}\left(1-\frac{\ell ^2-x^2+\tau ^2}{z^2}\right),
\end{align}
for which the induced metric becomes \eqref{eq:poincaremetric}.
Such coordinates are naturally associated with the conformal frame in which the boundary is 1+1 Minkowski space with line element $-d\tau^2+dx^2$. As stated in the main text, we take our global and Poincare coordinates to be related by
\eqref{eq:tant}, \eqref{eq:tanphi}, and \eqref{eq:r}.
In particular, with out conventions the line $\tau =0=x$ corresponds to $t=0, \phi=- \pi/2$ while the point $(\tau,x,z) = (+\infty, 0,0)$ is $(t, \phi, r) = (\pi\ell, -\frac{\pi}{2}, +\infty)$.
\begin{figure}[t]
\centerline{
\includegraphics[width=0.5\textwidth]{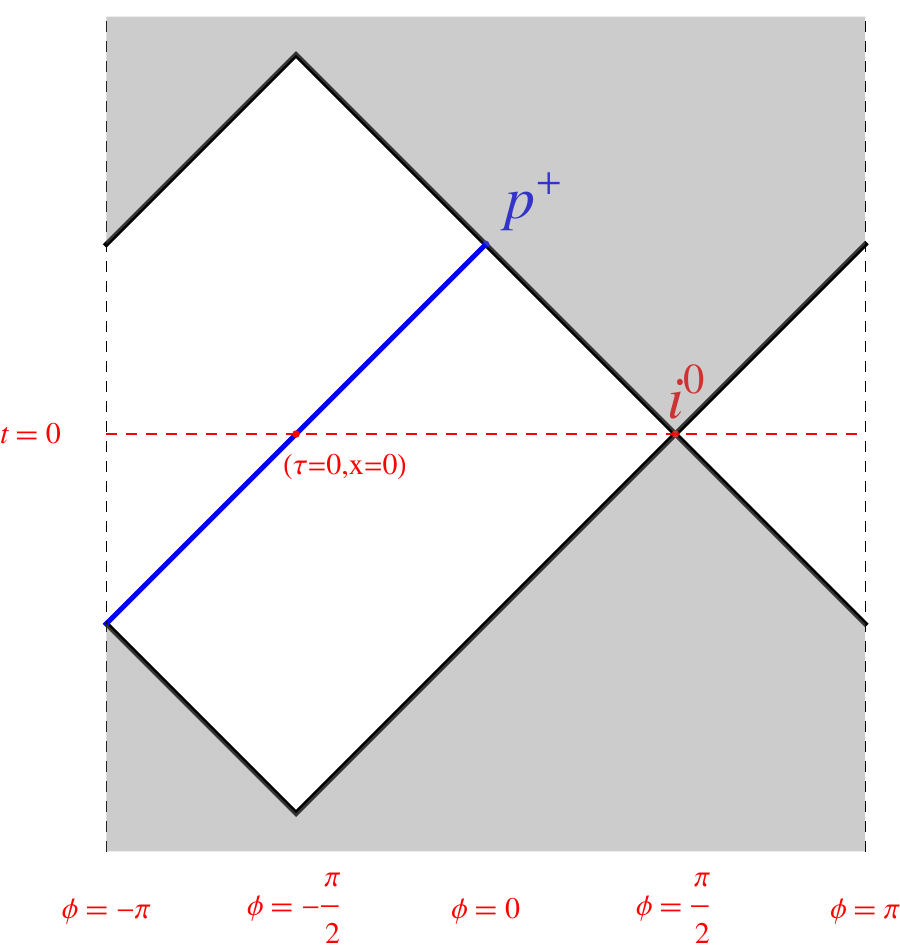}
}
\caption{The $S^1 \times {\mathbb R}$ boundary of AdS$_3$ shown here with our conventions for Poincar\'e and global coordinates. Our Poincar\'e patch (unshaded region) is not centered on the origin of our global coordinates but is instead displaced by a $\pi/2$ rotation. We have placed $p^+$ at $(t, \phi) = (\pi \ell/2,0)$ as in the main discussion of section \ref{sec:testing}.}
\label{fig:conventions}
\end{figure}

In the ambient 2+2 Minkowski space, one may consider the future Rindler Killing horizon $X^1 = T^1$ associated with the boost symmetry in the $T^1X^1$ plane.  The intersection of this surface with the hyperboloid \eqref{AdShyp} gives a future Rindler horizon in AdS$_3$.  In particular, labelling the individual null generators of this AdS$_3$ horizon by a parameter $\eta$, these generators satisfy
\begin{equation}
 X_2=\ell \sinh (\eta/\ell), \
 X_1=T_2, \  \text{and} \
 T_1= \ell \cosh(\eta/\ell) .
 \label{eq:22gen}
\end{equation}
We have chosen $\eta$ so that the induced (degenerate) line element on the horizon is just
\begin{equation}
ds^2  = d\eta^2;
\end{equation}
i.e., $\eta$ measures proper distance along the horizon. In Poincare coordinates, this horizon is just the surface $x=\tau$ with each generator lying at constant $z$; see figure \ref{fig:conventions}.  In particular, we note that there is no flux of horizon area though any $z=constant$ surface.

In terms of the global coordinates \eqref{eq:globalcoords1} and \eqref{eq:globalcoords2}, the relations \eqref{eq:22gen} take the form \eqref{eq:globHor}.
Taking $\eta \rightarrow \infty$ gives the generator along the conformal boundary corresponding to the 1+1 Rindler horizon.  Again using \eqref{eq:tant}, \eqref{eq:tanphi}, and \eqref{eq:r}, this generator satisfies \eqref{eq:tau}
for $\frac{t}{\ell }\in (-\frac{\pi }{2},\frac{\pi }{2})$, $\phi \in (-\pi,\pi )$.

\section{More General Bulk Rindler Horizons}
\label{sec:Append_GeneralHorizons}

In the main text, we considered only the case in which the point mass is motionless at the center of the spacetime and $p$ lies on the null curve $x = \tau$ as expressed in the GmD frame.  This case is not generic, as may be seen by recalling that the bulk horizon (traced backward from $p^+$) reaches the $r=0$ point mass at precisely $t(p)=0$.  Since $t(i^0) =0$, we see that this configuration is symmetric under the time reflection $t \rightarrow -t$.  But this symmetry is not shared by other configurations of $p, i^0$ (or equivalently of $p^+,p^-$) for which the CHI surface intersects $r=0$.  One might say that the case discussed in the main text has a defect that meets the CHI surface while ``at rest,'' while more general cases can feature a relative boost.

For completeness, we now compute $Q_{PmD}$ for arbitrary $p^+, p^-$.  This is merely a matter of acting on \eqref{eq:globHor} with $t, \phi$ translations to move $p^+$ into general position and manipulating the results in the same way as in the main text. We now find non-trivial results for any $\Gamma < 2 \pi$.  We comment on some key steps below.

As before, we proceed in two steps. But to simplify the calculations we introduce the monotonically increasing function $T=\tan \left(\frac{t}{\ell}+\phi _+\right)$ of the GmD time coordinate $t$.  We first find $dL_c/dT$ and then calculate $dT/d\lambda $.  Without loss of generality, we keep the point $i^0$ at $(t,\phi)=(0,\frac{\pi}{2})$ and we consider a general point $p^+$ to the future of $i^0$ along the left-moving null ray and thus lying at some $(t,\phi)=(\frac{\pi}{2}-\phi _+,\phi _+) $. Fixing $p^+$ and letting $p^-$ vary along the associated past-directed left-moving null ray defines points $p$ that satisfy $\frac{t}{l}=\phi +\frac{\pi}{2}-2\phi _+$ in GmD coordinates, or equivalently $\tau =x-\ell \tan \phi _+$ in PmD coordinates. We choose $\lambda =\tau (p)$ as the affine parameter along this null ray in the PmD frame. Setting $z=0$ in \eqref{eq:tanphi} yields
\begin{equation}
\label{generaltanphip}
\tan \phi _p=\frac{-\ell ^2+\left(\lambda +\ell \tan \phi _+\right)^2-\lambda ^2}{2\ell \left(\lambda +\ell \tan \phi _+\right)}.
\end{equation}
For $\phi _p \in \left(-\frac{\pi }{2},\phi _+\right)$, there is no intersection between the CHI surface and the defect; the area of CHI surface does not change. When $\phi _p=\phi _{p\text{ critical}}=-\frac{\pi }{2}$, the CHI surface first intersects the defect and $\lambda = \lambda _\text{critical}=: -\ell \tan \phi _+$.

It will again be convenient to take the defect to be diametrically opposed to $p^+$. For the general case we consider here, the defect is then $\phi \in \left(-\pi +\phi _+-\frac{\Gamma }{2},-\pi +\phi _++\frac{\Gamma }{2}\right)$.   But this imposes two restrictions.

First, it should not be possible to connect $i^0$ and $p^+$ by a timelike curve.  This requires $-\frac{\pi }{2}+\frac{\Gamma }{2}<\phi _+<\frac{\pi }{2}$ and is equivalent to keeping $i^0$ out of the defect.  Note that for fixed $\Gamma$ this condition allows the full (i.e., without defect) global coordinate $\bar \phi_+$ to to range over an interval of size $\pi$; i.e., as desired it allows all $p^+$ between $i^0$ and future timelike infinity of the original Poincar\'e patch on the boundary.  So this condition does not restrict the desired cases.  Instead, it is merely the transcription to GmD coordinates of the set of $p^+$ that we wish to study.

The second restriction comes from the fact that the
case $\Gamma  \ge \pi -2 \phi _+$ turns out to be trivial, as when tracing the generators backward from $p^+$ one finds that the ones along
generators along the boundary reach the defect before those in the interior as measured by the GmD time $t$.  For $\phi_+ =0$, this occurs for $\Gamma \ge \pi$ as one can see from figure \ref{fig:GmD} (right) using familiar causal properties of empty AdS$_3$.
 One can then show that the same is true as measured by our PmD $\lambda$. But since the length diverges near the boundary, this means that $L_c$ (defined by integrating $dL_c/dt$) is constant when there is no caustic and has an infinite discontinuity when the caustic forms.  This case is too singular to allow a meaningful analysis of \eqref{2dQNEC}, though \eqref{2dQNEC} is certainly consistent with simple regularizations of this result. We therefore concentrate on the remaining case $\phi _+<\frac{\pi }{2}-\frac{\Gamma }{2}$.  This second restriction is equivalent to keeping $p$ out of the defect.

For $\lambda <-\ell \tan \phi _+$, we have $dL_c/dT=2d\eta /dT$. The relevant bulk Rindler horizon (the past light cone of $p^+$) satisfies
\begin{equation}
\label{generalHor}
\left(T_2-X_1\right)\cos \phi _++(T_1-X_2)\sin \phi _+=0
\end{equation}
in terms of the embedding coordinates of appendix \ref{AdS3rev}.
Both the defect and the horizon we consider here can be obtained by acting with a rotation and a time translation on the defect and horizon studied in section \ref{sec:testing}. The intersection between the horizon and the defect thus satisfies the following relation obtained by acting in this way on equation \eqref{HorIntDef}:
\begin{equation}
\tanh \frac{\eta }{\ell}=T\tan \frac{\Gamma}{2}.
\end{equation}
Differentiating, we obtain
\begin{equation}
\label{generaldetadT}
\frac{d\eta }{dT}=\frac{\ell \tan \frac{\Gamma}{2}}{1-T^2\tan ^2\frac{\Gamma}{2}}.
\end{equation}

To find the remaining factor $dT/d\lambda $, we must locate the CHI surface.  This surface is the intersection between the past light cone of $p^+$ \eqref{generalHor} and the future light cone of $p^-$ defined by
\begin{equation}
\left(T_1-X_2\right)\cos \left(\phi _+-\phi _p\right)-\left(T_2+X_1\right)\sin \left(\phi _+-\phi _p\right)=0.
\end{equation}
We find
\begin{align}
\label{generalCHI}
\tan \frac{t}{\ell }=\frac{\cos \left(2\phi_+-\phi_p\right)\cos \phi}{\sin \left(\phi -\phi _p\right)+\sin \left(2\phi _+-\phi _p\right)\cos \phi },\\
r=\frac{\ell \cos \phi _p}{2\sqrt{\cos \phi \cos \phi _+\sin \left(\phi -\phi _p\right)\sin \left(\phi _+-\phi _p\right)}}.
\end{align}
Equation \eqref{generalCHI} further implies
\begin{equation}
T=\frac{\cos \left(\phi _+-\phi \right)}{-2\cos \phi \cos \phi _+\tan \phi _p+\sin \left(\phi_++\phi \right)}.
\end{equation}
After substituting \eqref{generaltanphip} into the above equation and setting $\phi =-\pi +\phi _++\frac{\Gamma}{2}$, differentiation yields
\begin{equation}
\label{generaldTdlambda}
\frac{dT}{d\lambda }=\frac{\cos \phi _++\cos \left(\phi _++\Gamma \right)}{2\ell \cos \phi _+\left(\cos \frac{\Gamma}{2}+\frac{\lambda}{l}\sin \frac{\Gamma}{2}\right)^2}.
\end{equation}

Combining \eqref{generaldetadT} and \eqref{generaldTdlambda}, we finally obtain
\begin{equation}
\frac{dL_c}{d\lambda }=2\frac{d\eta }{dT}\frac{dT}{d\lambda }=\frac{1}{\frac{\lambda }{\ell }+\frac{1}{2}\left(\cot \frac{\Gamma}{2}+\tan \phi _+\right)},\text{ for } \lambda <-\ell \tan \phi _+.
\end{equation}
Since $\frac{dL_c}{d\lambda }$ vanishes for $\lambda >-\ell \tan \phi _+$, a short computation shows our result to satisfy \eqref{eq:general}.  Thus our coarse-grained QNEC holds for general $p^+,p^-$.

\bibliographystyle{jhep}
	\cleardoublepage
\phantomsection
\renewcommand*{\bibname}{References}

\bibliography{QNEC}

\end{document}